\documentclass[aip,reprint,superscriptaddress]{revtex4-1}
\usepackage{amsmath}
\usepackage{amssymb}
\usepackage{dcolumn}
\usepackage{graphicx}
\usepackage{bm}
\usepackage[utf8]{inputenc}
\usepackage[T1]{fontenc}
\usepackage{mathptmx}

\begin{document}

\title[Effects of polymers on the cavitating flow around a cylinder]{Effects of polymers on the cavitating flow around a cylinder: A Large-scale molecular dynamics analysis} 



\author{Yuta Asano}
\email{yuta.asano@issp.u-tokyo.ac.jp}
\affiliation{Institute for Solid State Physics, The University of Tokyo, Kashiwa, Chiba 277-8581, Japan}

\author{Hiroshi Watanabe}
\affiliation{Department of Applied Physics and Physico-Informatics, Keio University, Yokohama, Kanagawa 223-8522, Japan}

\author{Hiroshi Noguchi}
\affiliation{Institute for Solid State Physics, The University of Tokyo, Kashiwa, Chiba 277-8581, Japan}


\date{\today}

\begin{abstract}
The cavitation flow of linear-polymer solutions around a cylinder is studied by performing a large-scale molecular dynamics simulation. The addition of polymer chains remarkably suppresses cavitation. The polymers are stretched into a linear shape near the cylinder and entrained in the vortex behind the cylinder. As the polymers stretch, the elongational viscosity increases, which suppresses the vortex formation. Furthermore, the polymers exhibit an entropic elasticity owing to stretching. This elastic energy increases the local temperature, which inhibits the cavitation inception. These effects of polymers result in the dramatic suppression of cavitation.
\end{abstract}


\maketitle 

%
%
\section{Introduction}\label{intro}
Cavitation represents a flow phenomenon in which bubbles are formed owing to the local pressure drop in a liquid~\cite{brennen2014}. The occurrence of cavitation adversely influences fluid machinery (pumps, turbines, etc.), in the form of performance degradation, noise, vibration, and erosion, among other effects. Notably, the generation of noise, vibration, and erosion may cause fatal damage to fluid machinery. Therefore, in engineering applications, it is vital to understand the cavitation mechanism and appropriately control the flow. In this context, the location of cavitation and its effects on the fluid machinery must be accurately predicted to design safe and highly efficient fluid machinery. In the cavitation process, tiny bubbles suddenly grow into large ones in a low-pressure region. In the primary bubble growth process, the interaction between the vortex and the bubble nucleus is considered to be of significance~\cite{arndt2002}. However, most industrial flows are turbulent, and it is challenging to predict the onset of cavitation in such irregular flows. Furthermore, flow with cavitation referred to as cavitating flow is highly complex as numerous bubbles repeatedly generate, grow, coalesce, disintegrate, and disappear. Therefore, analyzing the cavitating flow characteristics is a key issue in fluid engineering. Many researchers have attempted to examine the effects of cavitation on the flow field and fluid machinery and develop cavitation models~\cite{ks2010, yda2020, hzg2016, sap2017, wwh2017, uk2018, apb2015, gjn2020}.

The realization of flow control by adding ingredients has attracted attention, as the performance and durability of fluid machinery can be enhanced without changing the machinery design. In particular, the addition of a tiny amount of polymer to cavitating flows can suppress the onset of cavitation~\cite{ewt1970, ting1978}. Moreover, polymer addition can help suppress the noise, vibration, and erosion caused by cavitation~\cite{nst1986, tsujino1987, bhw2008, ryskin1990}. To understand the underlying mechanism, it is crucial to clarify the polymer behavior in the flow. Therefore, the effects of polymer addition in cavitation flows have been investigated for various engineering applications such as propellers, pumps, injectors, and water jets~\cite{ply2014, alr2019, fpc1995, nkm2018, ktk2018}. It has been reported that the increase in the elongational viscosity caused by the stretching of the polymer by the flow is of significance in suppressing the cavitation and mitigating the impact pressure generated during bubble collapse. Furthermore, because cavitation can occur in various forms depending on the flow conditions, the relationship between cavitation occurrence and viscoelasticity was experimentally and numerically investigated~\cite{nkm2018, ktk2018}. They studied the effects of polymer on cavitation flow in step and injector nozzles. In general, for the flow in the step nozzle, cloud cavitation caused by the detachment occurs predominantly because of the flow separation; in contrast, in the case of the injector nozzle, both string cavitation and cloud cavitation occur. The string cavitation can be attributed to the vortices formed in the nozzle suck. The addition of polymers suppresses cloud cavitation but enhances longitudinal vortex cavitation. A significant change in the vortex structure considerably influences the suppression and enhancement of cavitation. Therefore, to clarify the effects of polymer addition on cavitation, it is necessary to simultaneously consider the vortex, polymer motion, and phase transition dynamics in the flow.

Flow around a circular cylinder is a representative example of flows with vortices. Although such flows are simple, they involve many essential physical phenomena such as vortex motion~\cite{pcl1982, gerrard1966}, drag behavior~\cite{henderson1995}, and Aeolian sound~\cite{phillips1956}. Because the flow characteristics of Newtonian fluids are well understood~\cite{williamson1996}, the effects of polymer additives~\cite{cbg2001, so2004, ris2010, nvm2013, xby2017, pm2006, cp2003} and cavitation~\cite{kcb2017, fry1984, sms2008, gm2016, bbm2020} have extensively been examined. Reitzer {\it et al.} experimentally investigated the effects of polymer addition on the cavitation flow around a cylinder~\cite{rgs1985}. They measured the sound pressure generated by cavitation in water and polymer solutions. The results indicated that the polymers suppressed the onset of cavitation and reduced the drag force and noise caused by cavitation. Moreover, with the increase in the degree of cavitation, no significant difference was observed between the water and polymer solutions. They concluded that these effects were related to the viscoelasticity of the polymer solution and elongation properties of the polymer. To further understand the cavitation suppression mechanism, it is important to clarify the elongation properties of polymers in cavitation flow from a molecular viewpoint.

In this study, the cavitation flow around a circular cylinder with polymers is investigated from a molecular scale by performing molecular dynamics (MD) simulations. Previously, we examined the effects of polymer addition~\cite{awn2018} and cavitation~\cite{awn2020} on flow around cylinders. Here, combining these two conditions, we simulated the cavitation in polymer solutions and clarified the effects of polymers by comparing the results with those of a Newtonian fluid. The micro-scale insights obtained through MD calculations can facilitate the understanding of the suppression mechanism of cavitation through polymer addition and detailed modeling of cavitation and polymers. The remaining paper is organized as follows: Sec. II describes the simulation model and methodology. The results are presented in Sec. III and discussed in Sec. IV. Section V presents the concluding remarks.
\section{Method}\label{method}
A solvent particle is a monatomic molecule, and its interparticle interaction is the smoothed-cutoff Lennard-Jones potential function~\cite{td2011}
\begin{eqnarray}
  \label{eq:sLJ}
  u_{\rm sLJ}(r;r_{\rm c})&=&
  \left\{
  \begin{array}{ll}
    u_{\rm LJ}(r) - u_{\rm LJ}(r_{\rm c}) - (r-r_{\rm c})u'_{\rm LJ}(r_{\rm c}) & (r\le r_{\rm c}) \\
    0                                                                           & (r>r_{\rm c})
  \end{array} \right.,\\
  u_{\rm LJ}(r)&=&4\epsilon\left[\left(\frac{\sigma}{r}\right)^{12} - \left(\frac{\sigma}{r}\right)^{6}\right],
  \label{eq:LJ}
\end{eqnarray}
where $r$ is the interparticle distance, and $\epsilon$ and $\sigma$ denote the energy and length scales, respectively. $r_{\rm c}$ is the cutoff distance of the potential function, with $r_{\rm c}=2.5\sigma$ for the solvent-solvent interaction. The prime in Eq.~(\ref{eq:sLJ}) represents the derivative to $r$.

We consider a linear polymer chain and adopt the Kremer--Grest model for its interaction~\cite{kg1990}. The interactions between the polymer beads and solvent particles and those between the polymer beads are defined as $u_{\rm sLJ}(r;2.5\sigma)$ and $u_{\rm sLJ}(r;2^{1/6}\sigma)$, respectively, except for the nearest neighbor beads along the chain. These nearest neighbor beads are connected by a bond by using a FENE (finitely extensible nonlinear elastic) potential:
\begin{eqnarray}
  u_{\rm bond}(r) &=& u_{\rm sLJ}(r;2^{\frac{1}{6}}\sigma)+u_{\rm FENE}(r), \label{eq:bond}\\
  u_{\rm FENE}(r)&=&
  \left\{
  \begin{array}{ll}
    -\frac{K}{2}R_{0}^{2}\ln \left[1-\left(\frac{r}{R_0}\right)^2\right] & (r\le R_{0}) \\
    \infty                                                             & (r>R_{0})
  \end{array}
  \right.
  ,\label{eq:FENE}
\end{eqnarray}
where $K$ and $R_{0}$ denote the strength of the interaction and equilibrium length between the beads, respectively. Here, we set $K = 30 \epsilon/\sigma^2$ and $R_{0} = 1.5 \sigma$. The masses of the solvent particle and polymer bead are $m$. All the physical quantities are represented in the units of energy $\epsilon$, length $\sigma$, and time $\sigma\sqrt{m/\epsilon}$. The fluid without polymer addition is referred to as an LJ fluid.

\begin{figure}
  \includegraphics{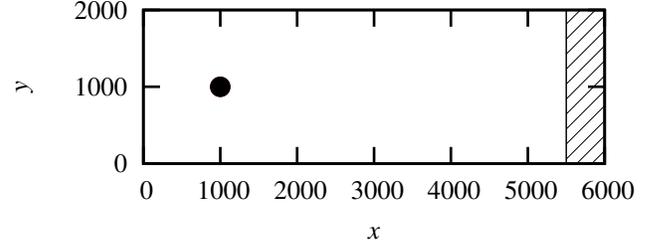}%
  \caption{\label{fig:box} Schematic of the simulation box. The black circle represents a circular cylinder of diameter $D=250$, and the central axis is located at $(x_{\rm c}, y_{\rm c})=(1000, 1000)$. The shaded area $(5500\le x \le 6000)$ is the region of equilibration and velocity control implemented by the Langevin thermostat for $T=T_{\rm in}$. The thickness in the $z$ direction is $100$.}%
\end{figure}

The simulation box is rectangular with dimensions $L_x \times L_y \times L_z =6000 \times 2000 \times 100$ (Fig.~\ref{fig:box}). The periodic boundary conditions are imposed in all directions. The filled circle in Fig.~\ref{fig:box} represents a circular cylinder, modeled by fixing particles on the surface of the cylinder. The interaction between a particle of the cylinder and a particle of the solution is defined as $u_{\rm sLJ}(r;2.5)$. The diameter of the cylinder is $D=250$. The cylinder axis is parallel to the $z$-direction and its position is $(x_{\rm c}, y_{\rm c})=(1000, 1000)$. A Langevin thermostat for the temperature $T=T_{\rm in}$ of the input flow is imposed in the shaded area $5500 \le x \le 6000$ in Fig.~\ref{fig:box} for equilibration and velocity control for the fluid. The friction coefficient of the Langevin thermostat is increased linearly with $x$ from $0.0001$ to $0.1$ for $5500<x<5750$ and set as $0.1$ in the remaining region. The velocity of the fluid is set as $V = 1$ in the $x$-direction. This type of a local Langevin thermostat has previously been used to examine the K\'arm\'an vortex~\cite{awn2018,awn2019,awn2020} and sound wave~\cite{awn2020a}.

As an initial setup, fluid particles are randomly placed in the simulation box while avoiding particle overlaps, except inside the cylinder. The initial velocities of the fluid particles are generated according to the Maxwell velocity distribution with $T=T_{\rm in}$ and average velocity $V$ in the $x$-direction.  The total number of particles in the fluid is $N=N_{\rm LJ}+N_{\rm s}N_{\rm p}$, where $N_{\rm LJ}$ is the number of solvent particles and $N_{\rm s}$ and $N_{\rm p}$ denote the number of polymer beads and the number of polymers, respectively. In this study, we set $N_{\rm s}=500$. The polymer concentration is the overlap concentration $c^*=0.0069$ obtained from the Flory theory~\cite{bgm2013}. We study the two types of fluids, the LJ fluid with $N_{\rm p} = 0$ and polymer solution with $N_{\rm p} = 16~567$ ($c=c^*$), to investigate the polymer effects. Moreover, to clarify the effects on cavitation, simulations are performed at two temperatures $T_{\rm in}=2$ and $T_{\rm in}=1.25$ (temperatures are given with the energy units, i.e., with the Boltzmann constant $k_{\rm B}=1$). With reference to our previous study~\cite{awn2020}, for the LJ fluid, $T_{\rm in}=2$ and $T_{\rm in}=1.25$ represent a non-cavitating flow and cavitating flow, respectively. The density of the fluid is fixed as $\rho = N/L_xL_yL_z = 0.4$ for the flow simulations ($N=478~036~504$) and for the calculation of the gas--liquid phase boundary and viscosity as described later in this section.
Following Ref.~\onlinecite{awn2020}, the density of the fluid is fixed as for cavitation flow around a circular-cylinder array of LJ fluid~\cite{awn2020}.

LAMMPS (Large-scale Atomic/Molecular Massively Parallel Simulator)~\cite{primpton1995} is used to perform the numerical integration of the equation of motion. The velocity Verlet algorithm is used for the time integration. The numerical integration is performed for about $4~000~000$ steps with a time step of $0.004$. Because the steady-state of vortex shedding is achieved at $3~000~000$ steps, the data for the subsequent $1~000~000$ steps are used for the analysis. The statistical averages and errors are determined from three or more simulations with different initial configurations and initial velocities. For flow simulations, $144$ nodes of the ISSP supercomputer (AMD EPYC 7702, 64 cores$\times 2$ per node) are used. One simulation is parallelized using the flat-message passing interface (MPI) ($18~432$ MPI processes), and its typical execution time is $\sim 100$ h.

\begin{figure}
  \includegraphics[width=3.37in]{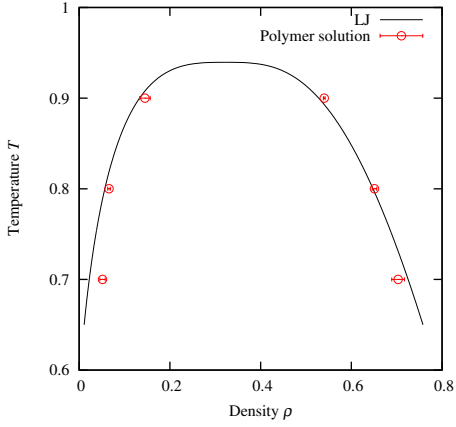}%
  \caption{\label{fig:phaseDiagram} Gas--liquid phase boundary for the polymer solution (circles) and LJ fluid~\cite{awn2020} (line). The numbers of polymer beads and mean polymer concentration are $500$ and $0.0069$, respectively.}%
\end{figure}

We first estimate the gas--liquid phase boundary without the flow. Specifically, the gas--liquid phase boundary of the polymer solution at equilibrium is estimated through gas--liquid coexistence simulations~\cite{wih2012}. The system is a rectangle with dimensions of $L_x \times L_y \times L_z = 1000 \times 250 \times 250$. Periodic boundary conditions are imposed in all directions. At the initial step, the particles are placed in a liquid state for $x<250$ and in a gas state for the remaining system. The Langevin thermostat is applied to control the temperature of all the particles. The liquid and gas densities in the gas--liquid coexistence state are estimated as a function of temperature. The number of polymer beads and the concentration of polymers are equal to those in the flow simulation. The black line in Fig.~\ref{fig:phaseDiagram} shows the gas--liquid phase boundary of the LJ fluid obtained in our previous simulation~\cite{awn2020}. In the absence of flow, the effects of the polymers are negligible, as shown in Fig.~\ref{fig:phaseDiagram}.

\begin{figure}
  \includegraphics{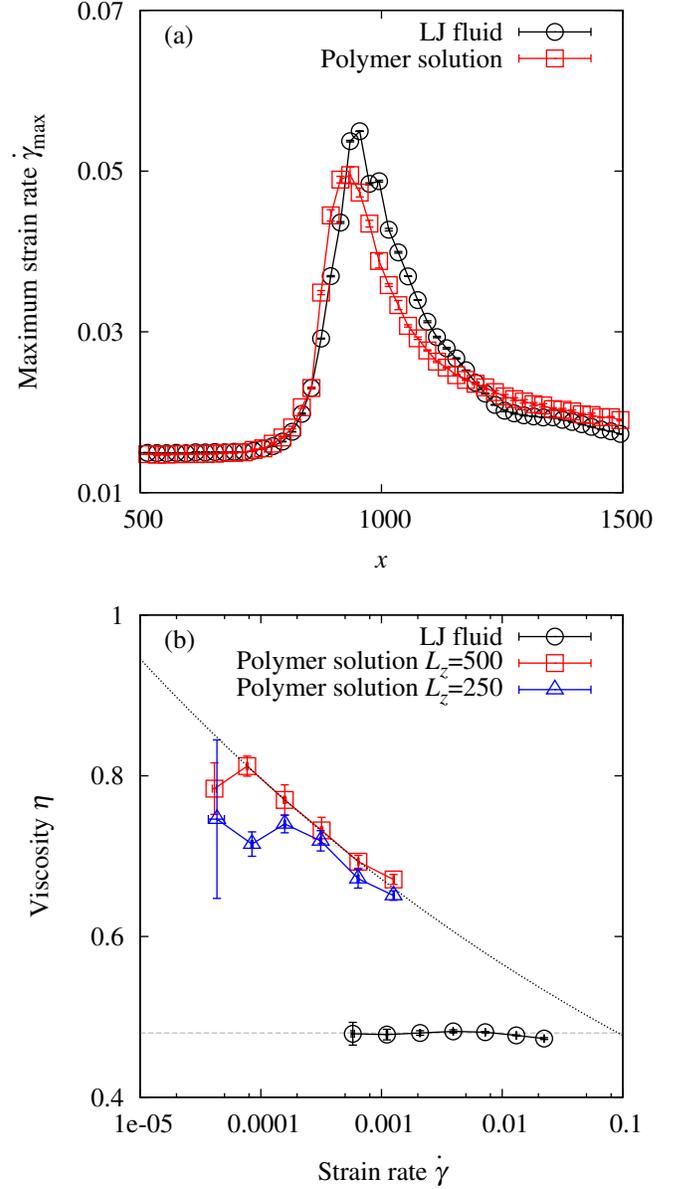}%
  \caption{\label{fig:viscosity} (a) Maximum shear rate $\dot{\gamma}_{\rm max}$ as a function of $x$ for the flow around the cylinder at temperature $T_{\rm in}=2$. (b) Shear rate $\dot{\gamma}$ dependence of the viscosity $\eta$ for the LJ fluid and polymer solution at $T=2$. The dotted line shows the power function obtained by fitting the shear-thinning part $(0.0001 \lesssim \dot{\gamma} \lesssim 0.002)$ of the polymer solution. The dashed line shows the viscosity of the LJ fluid, $\eta=0.48$.}%
\end{figure}

Next, we estimate the viscosity $\eta$ to calculate the Reynolds number $Re=\rho D V / \eta$, which characterizes the flow. First, we determine the characteristic shear rate of the system. Figure~\ref{fig:viscosity}(a) shows the maximum shear rate $\dot{\gamma}_{\rm max}$ at each position $x$ for the flow around the cylinder at temperature $T_{\rm in}=2$ (compare it with Fig.~\ref{fig:box}). The shear rate is estimated as the difference between the maximum and minimum eigenvalues of the velocity gradient tensor. The shear rate exhibits the maximum at the vicinity of the cylinder. Thus, this maximum shear rate, $\dot{\gamma}=0.05$, is chosen as the characteristic shear rate.
The viscosities $\eta$ for the LJ fluid and polymer solution are estimated by generating two Couette flows with positive and negative shear rates, by using the momentum exchange method~\cite{muller99}. The details of $\eta$ calculation are described in the Appendix. Figure~\ref{fig:viscosity}(b) shows the  $\eta$ dependence on $\dot{\gamma}$. The viscosity of the LJ fluid exhibits nearly no dependence on the shear rate. In contrast, the viscosity of the polymer solution shows shear-thinning behavior in the range of shear rate $0.0001 \lesssim \dot{\gamma} \lesssim 0.002$. The shear-thinning behavior is almost independent of the finite size effects. However, the viscosity estimation for higher shear rates, $\dot{\gamma}\gtrsim 0.006$, is difficult due to the finite size effects (see the Appendix).
Although a much larger system is required to estimate the viscosity at high shear rates, however, the calculation for such a large system size is not feasible given the limitation of the available computational resources. Alternatively, we estimate the viscosity at $\dot{\gamma}=0.05$ through extrapolation and fit the shear-thinning part $(0.0001 \lesssim \dot{\gamma} \lesssim 0.002)$ by using a power function.  The dotted line in Fig.~\ref{fig:viscosity}(b) shows the power function, $\eta=0.40\dot{\gamma}^{-0.074}$.
The exponent of shear rate is similar to that of the results using multiparticle collision dynamics simulation~\cite{ry06}.
The viscosity coefficient of the polymer solution at $\dot{\gamma} = 0.05$ is approximately the same as that of the LJ fluid. Therefore, in this study, the polymer solution and LJ fluid have the same Reynolds number under the same inflow conditions of the density, temperature, and flow velocity. Thus,  $\eta= 0.48$  and $0.40$ are used for $T_{\rm in}=2$ and $1.25$, respectively, and the Reynolds numbers of the system are $Re=210$ and $240$ for  $T_{\rm in}=2$ and $1.25$, respectively, which are in the range of the formation of the K\'arm\'an vortex.
Moreover, we estimate the rheological properties of the flow by the Weissenberg number $Wi=\tau_{\rm r}\dot{\gamma}$ ($\tau_{\rm r}$ is the polymer relaxation time). The relaxation time is estimated by the Zimm relaxation time as $\tau_{\rm r} \simeq \eta_{\rm s} N_{\rm s}^{1.8}/T \simeq 20~000$, where $\eta_{\rm s}$ is the solvent viscosity~\cite{hgr13, de88}. Because of $Wi\simeq 1000 \gg 1$ at $\dot{\gamma}=0.05$, the flow is a strongly nonlinear region.

The vorticity, density, and temperature fields are calculated by dividing the simulation cell into cubic grids with a side length of $10$. Each grid is treated as a fluid element. The vorticity $\omega_{z}$, density $\rho$, and temperature $T$ of each fluid element are calculated using the following equations:
\begin{eqnarray}
  \omega_z&=&\frac{\partial v_y}{\partial x} - \frac{\partial v_x}{\partial y},\label{eq:omega_z}\\
  \rho&=&\frac{n}{V_{\rm grid}},\label{eq:rho}\\
  T&=&\frac{2}{3n}\left(K_{\rm total} - K_{\rm trans} \right),\label{eq:T}\\
  v_{k}&=&\frac{1}{n}\sum_{i} v_{i k},\label{eq:vk}\\
  K_{\rm total} &=& \sum_{i, k} \frac{v_{i k}^2}{2},\label{eq:Ktotal}\\
  K_{\rm trans} &=& n\sum_{k=x,y,z}\frac{v_{k}^2}{2},\label{eq:Ktrans}
\end{eqnarray}
where $k \in \{x,y,z\}$, $n$ is the number of fluid particles in the fluid element, and $V_{\rm grid}$ is the volume of the fluid element. $K_{\rm total}$ and $K_{\rm trans}$ represent the total kinetic energy and the kinetic energy of the center of mass of the fluid element, respectively. $v_{ik}$ is the $k$th component of the $i$th fluid particle velocity in the fluid element. The summations in Eqs.~(\ref{eq:vk}) and (\ref{eq:Ktotal}) are carried out for all the fluid particles in the fluid element. The derivative in Eq.~(\ref{eq:omega_z}) is calculated using the central difference formula using the flow velocities of the neighboring fluid elements.
\section{Results}\label{results}
\subsection{Non-cavitating flow}\label{highT}
\begin{figure*}
  \includegraphics{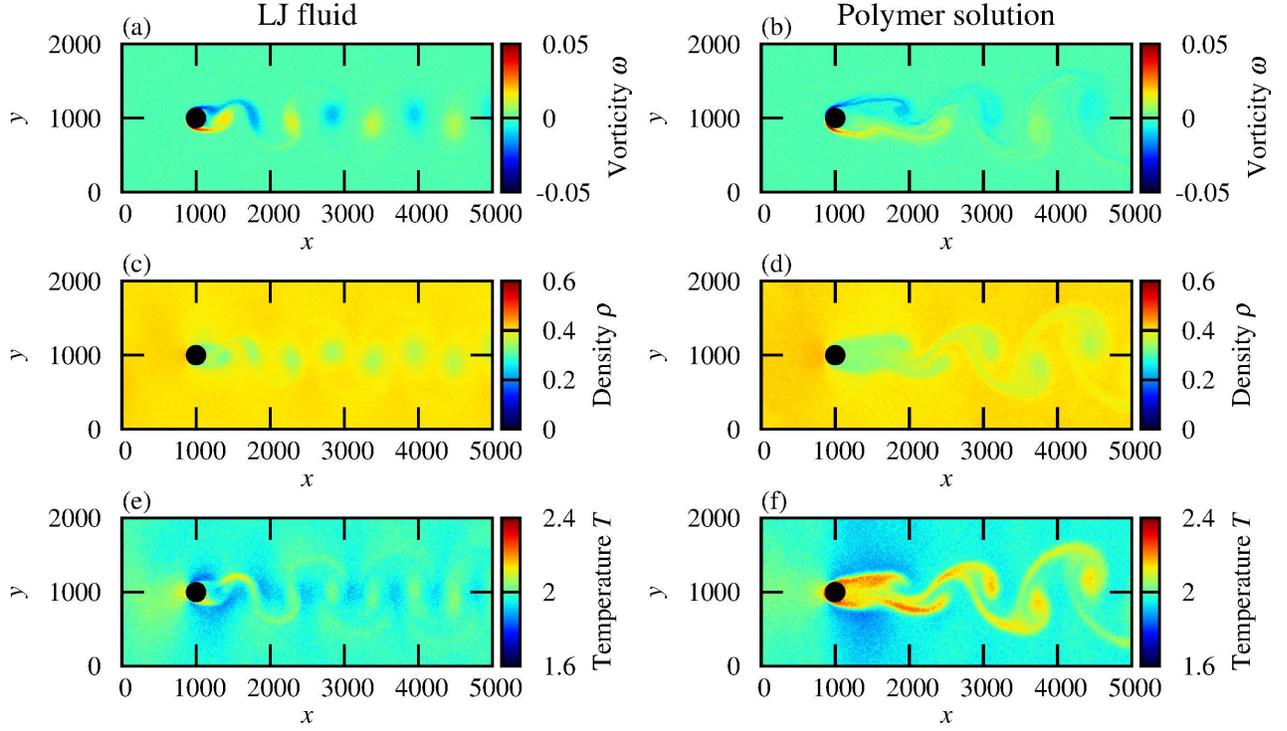}%
  \caption{\label{fig:highT} Snapshots of (a) vorticity field $\omega_z$, (c) density field $\rho$, and (e) temperature field $T$ of the LJ fluid and (b) $\omega_z$, (d) $\rho$, and (f) $T$ of the polymer solution at $T_{\rm in}=2$.}%
\end{figure*}

At a high temperature ($T_{\rm in}=2$), the polymer solution and LJ fluid do not exhibit any cavitation. The vorticity, density, and temperature fields are shown in Fig.~\ref{fig:highT}. 

For the LJ fluid, the staggered vortex street is generated behind the cylinder (Fig.~\ref{fig:highT}(a)), that is, the K\'arm\'an vortex. The density is low in the separated shear layer, recirculation region, and center of the vortex behind the cylinder (Fig.~\ref{fig:highT}(c)). The temperature is low in the separated shear layer and between the vortices behind the cylinder (Fig.~\ref{fig:highT}(e)). In contrast, in the case of the polymer solution, the recirculation region is extended, and the vortices behind the cylinder are blurred (Fig.~\ref{fig:highT}(b)).  Similar to the case of the LJ fluid, the density is low at the center of the vortex (Fig.~\ref{fig:highT}(d)). With the modification of the flow field, large regions of high temperature emerge behind the cylinder, unlike in the case of the LJ fluid. Furthermore, the temperature is low in the region outside the separated shear layer, similar to that in the case of the LJ fluid (Fig.~\ref{fig:highT}(f)).

These features of the polymer solution are in agreement with those observed in our previous simulation in two-dimensional (2D) space~\cite{awn2018} and those reported in other past simulation~\cite{so2004,ris2010,nvm2013,xby2017} and experimental studies~\cite{cbg2001,cp2003,pm2006}. For 3D simulation, a polymer length  $N_{\rm s}\gtrsim 500$ is required to obtain the blurring, whereas  $N_{\rm s}=100$ is sufficient for 2D simulation~\cite{awn2018}. Note that a slightly large vortex shedding frequency is obtained owing to the finite size effects reported in our previous study~\cite{awn2019}.

\subsection{Cavitating flow}\label{lowT}
\begin{figure*}
  \includegraphics{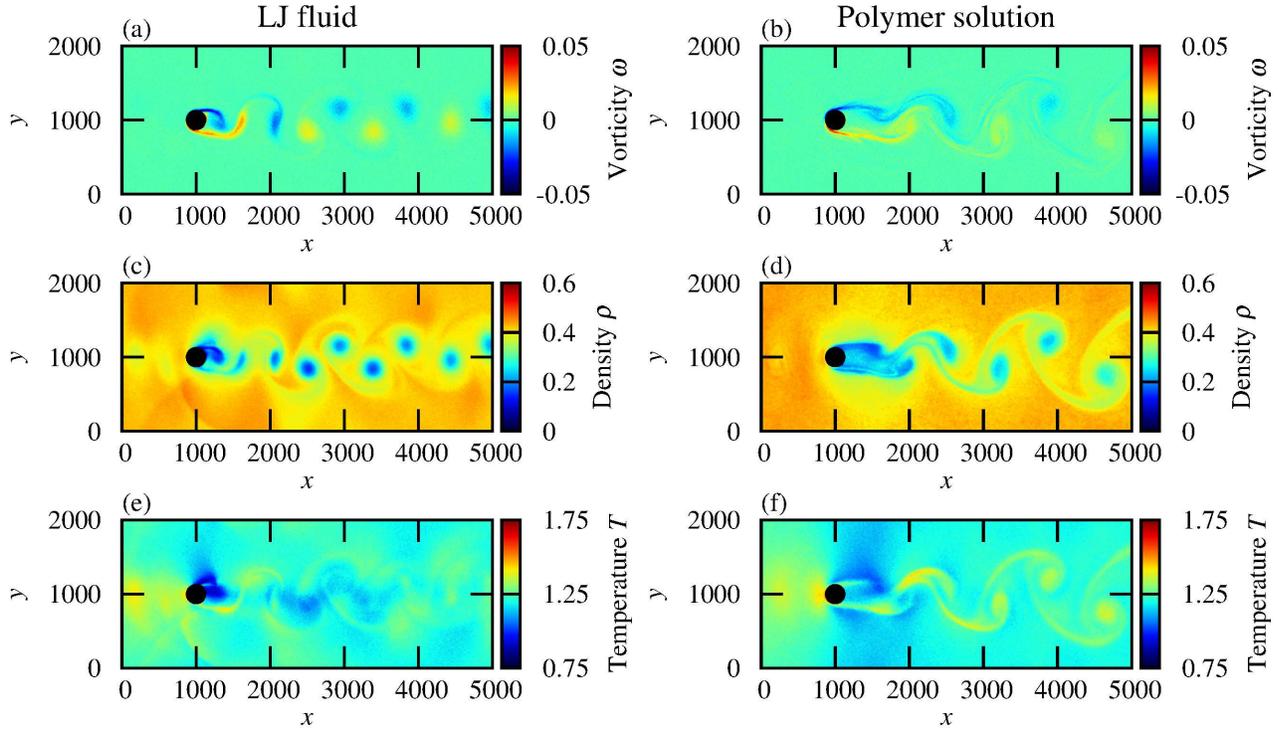}%
  \caption{\label{fig:lowT} Snapshots of (a) $\omega_z$, (c) $\rho$, and (e) $T$ of the LJ fluid and (b) $\omega_z$, (d) $\rho$, and (f) $T$ of the polymer solution at $T_{\rm in}=1.25$.}%
\end{figure*}

Figure~\ref{fig:lowT} shows the flow behaviors at a low temperature ($T_{\rm in}=1.25$), which pertains to the region of cavitation of the LJ fluid, as reported in our previous study~\cite{awn2020}. In the vorticity field, the distance between the upper and lower vortices is slightly larger for both fluids compared to that in the non-cavitation flow. The density becomes much lower in the center of the vortex than in non-cavitation flow (compare Figs.~\ref{fig:lowT}(c),(d) and \ref{fig:highT}(c),(d)). However, the amount of density decrease is slightly less in the polymer solution than that in the LJ fluid. On the other hand, the temperature increase in the wake region is largely reduced in the polymer solution (compare Figs.~\ref{fig:lowT}(f) and \ref{fig:highT}(f)). Similarly, the temperature decreases more for the LJ fluid  (compare Figs.~\ref{fig:lowT}(e) and \ref{fig:highT}(e)).

\begin{figure*}
  \includegraphics{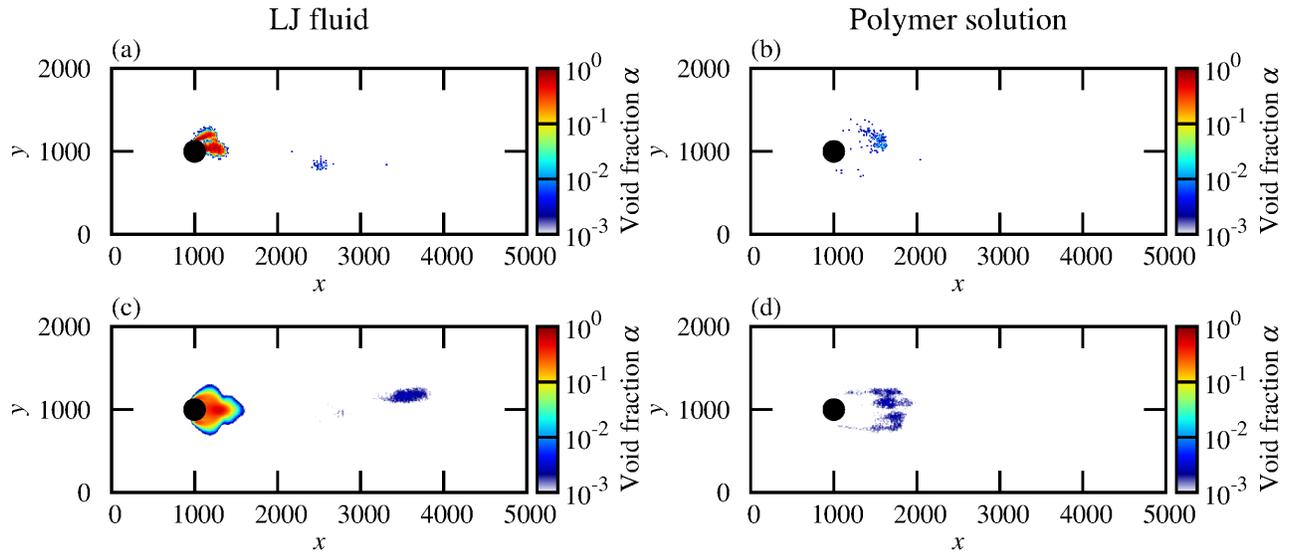}%
  \caption{\label{fig:void} Void fraction $\alpha$ for the (a) and (c) LJ fluid and (b) and (d) polymer solution at $T_{\rm in}=1.25$.
The instantaneous and time-averaged void fractions are shown in (a), (b), (c), and (d), respectively.}
\end{figure*}

We evaluate the void fraction $\alpha$ to examine the effects of the polymers on the onset of cavitation. $\alpha$ is obtained by assuming the local equilibrium for the fluid element at each time step, as follows:
\begin{eqnarray}
  \rho=\alpha \rho_{\rm gas}(T) + (1-\alpha)\rho_{\rm liq}(T),\label{eq:alpha}
\end{eqnarray}
where $\rho_{\rm gas}(T)$ and $\rho_{\rm liq}(T)$ denote the densities of the gas and liquid phases in the gas--liquid equilibrium state for the LJ fluid at $T$, respectively. The spatial distributions of the void fraction are shown in Fig.~\ref{fig:void}. For the LJ fluid, bubbles appear in the separated shear layer and recirculation region behind the cylinder. Moreover, bubbles are observed in the center of the vortices. In contrast, bubble generation is highly reduced for the polymer solution. Almost no bubbles are present in the separated shear layer and recirculation region; a few bubbles are observed near the point at which the vortex is emitted. Therefore, the addition of polymers suppresses the bubbles and changes the location of the bubble generation.

\begin{figure*}
  \includegraphics{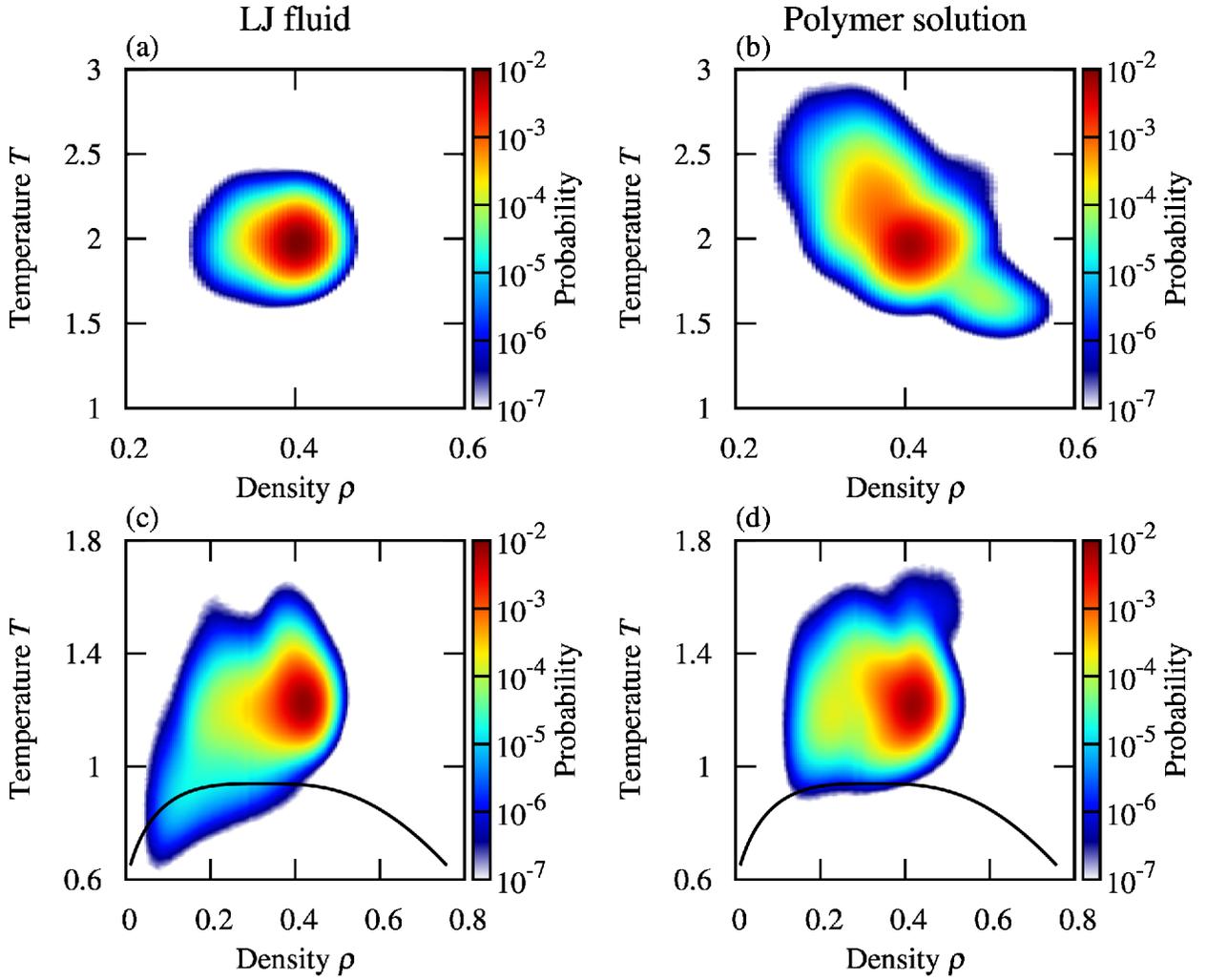}%
  \caption{\label{fig:map} Probability distributions of thermodynamic states of each fluid element for the (a) LJ fluid at $T_{\rm in}=2$, (b) polymer solution at $T_{\rm in}=2$, (c) LJ fluid at $T_{\rm in}=1.25$, and (d) polymer solution at $T_{\rm in}=1.25$. The black lines in (c) and (d) indicate the gas--liquid phase boundaries of the LJ fluid.}%
\end{figure*}

To understand the mechanism of this cavitation suppression, the thermodynamic states of the fluid elements are investigated. The distribution of the thermodynamic state on the temperature-density diagrams is shown in Fig.~\ref{fig:map}. Under the assumption of local equilibrium in each fluid element at each time step, the temperature and density are calculated using Eqs.~(\ref{eq:rho}) and (\ref{eq:T}), respectively. The thermodynamic states are classified into intervals of temperature and density width of $\Delta T = 0.01$ and $\Delta \rho = 0.01$, respectively. In the non-cavitating flow, the distribution of the LJ fluid is relatively symmetric around the inflow condition ($\rho=0.4, T_{\rm in}=2$). In contrast, the distribution of the polymer solution is asymmetric and broader. At a lower density, the local temperature is higher.  Interestingly, this temperature trend is more pronounced in the cavitating flow ($T_{\rm in}=1.25$). In the case of the LJ fluid, the thermodynamic state is distributed largely below the critical point (deeply in the phase separation region and low-temperature and low-density regions). In contrast, most of the thermodynamic states are in the supercritical region in the case of the polymer solution. Bubbles are generated below the critical point. Therefore, this increase in the local temperature is considered to be the origin of the cavitation suppression.

\begin{figure*}
  \includegraphics{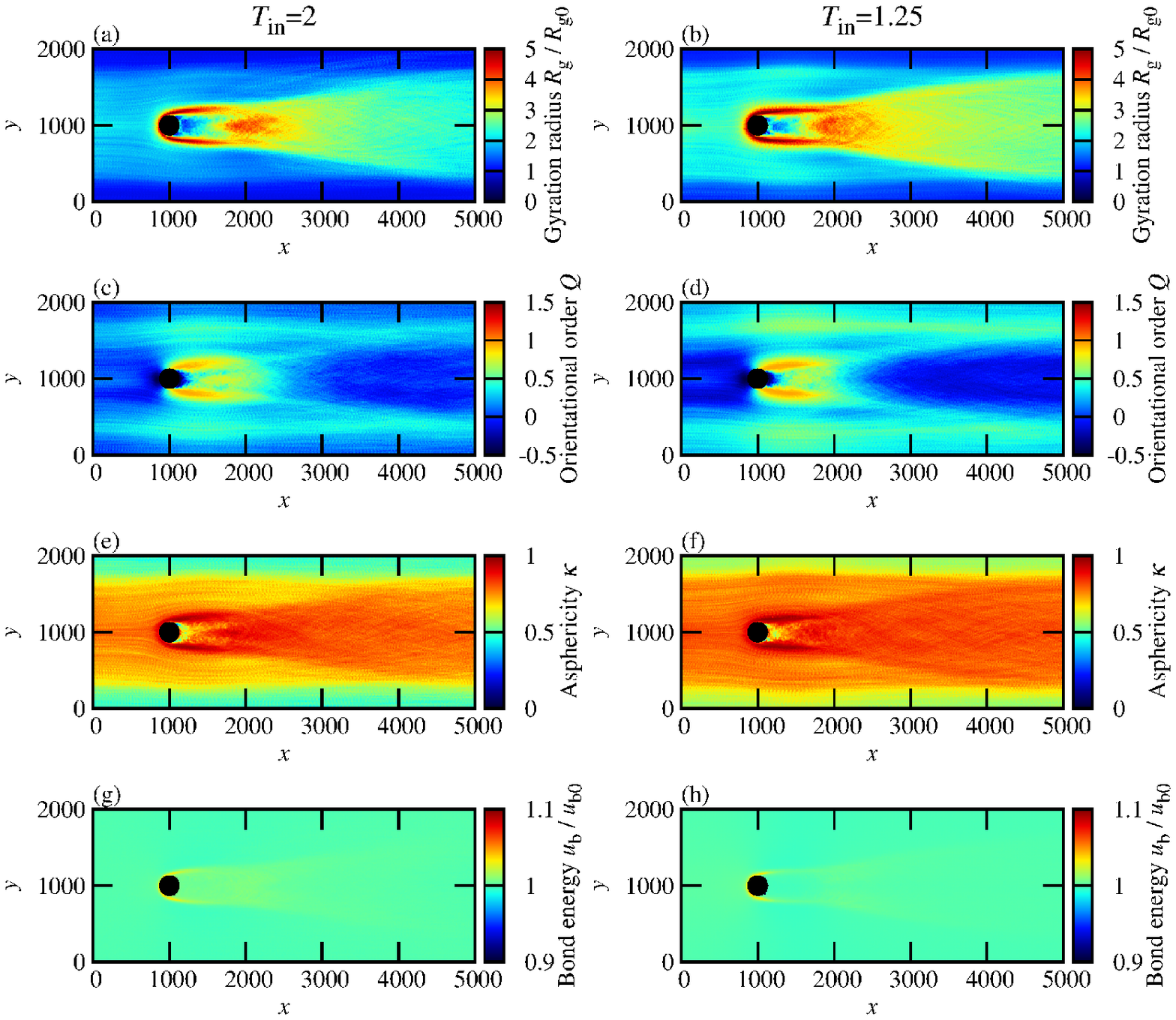}%
  \caption{\label{fig:polymer} Time-averaged polymer conformations at (a), (c), e), and (g) $T_{\rm in}=2$ and (b), (d, (f, and (g) $T_{\rm in}=1.25$. [(a) and (b)] Gyration radius $R_{\rm g}$. [(c) and (d)] Orientational order $Q$. [(e) and (f)] Asphericity $\kappa$. [(g) and (h)] Bond energy $u_{\rm b}$. $R_{\rm g0}$ and $u_{\rm b0}$ denote the equilibrium values of gyration radius and bond energy, respectively, for an isolated polymer chain in the absence of flow.}%
\end{figure*}

Next, we investigate the polymer conformation in the flow considering its influence on the thermodynamic states of the flows. The polymer chain is assumed to belong to the fluid element involving its center of mass ${\mathbf r}_{\rm G}$. At each time step, the gyration radius $R_{\rm g}$, orientation order $Q$, and asphericity $\kappa$ are calculated from the gyration tensor $S_{kl}$ of the polymer,
\begin{eqnarray}
  S_{k l} &=& \frac{1}{N_{\rm s}}\sum_{i=1}^{N_{\rm s}}r_{i k} r_{i l},\label{eq:smn}\\
  R_{\rm g}^2 &=& \lambda_1 + \lambda_2 + \lambda_3,\label{eq:rg}\\
  Q &=& \frac{1}{2} \left(3\cos^2 \theta - 1\right),\label{eq:q}\\
  \kappa &=& \frac{(\lambda_1-\lambda_2)^2 + (\lambda_2-\lambda_3)^2 + (\lambda_3-\lambda_1)^2}{2(\lambda_1 + \lambda_2 + \lambda_3)^2},\label{eq:kappa}
\end{eqnarray}
where $r_{ik}$ is the $k$th component of the relative position of the $i$th bead from ${\bm r}_{\rm G}$. $\lambda_{j}$ $(j=1, 2, 3)$ is an eigenvalue of the gyration tensor $S_{k l}$, where $\lambda_{1} \ge \lambda_{2} \ge \lambda_{3}$. $\theta$ is the angle between the eigenvector associated with the maximum eigenvalue $\lambda_{1}$ of the gyration tensor and the $x$-direction. Hence, $Q$ is the orientational order of the polymer in the flow direction: $Q=1$ when the polymer is oriented in the flow direction, $Q=-0.5$ when the polymer is oriented perpendicularly to the flow direction, and $Q=0$ when the polymer is randomly oriented. $\kappa$ indicates the deviation from a sphere~\cite{rudn86,nogu98,nogu05}: $\kappa=0$ when the polymer is spherical, and $\kappa=1$ when the polymer is completely stretched in one direction. The time-averaged conformation of the polymer in each fluid element is shown in Fig.~\ref{fig:polymer}. The polymer is stretched into a linear shape by the shear near the cylinder for both non-cavitating and cavitating flows. Moreover, the polymer is entrained in the vortex behind the cylinder. The polymers are more stretched at $T_{\rm in}=1.25$, but it is only a moderate change.  The bond energy between beads, $u_{\rm b}$, is virtually the same as the equilibrium value of an isolated polymer in the absence of flow.

\section{Discussion}\label{discussion}
As shown in Figs.~\ref{fig:void} and \ref{fig:map}, cavitation is significantly suppressed by the polymer addition, although the polymer has negligible effects on the gas--liquid phase equilibrium without flow (see Fig.~\ref{fig:phaseDiagram}). Therefore, the polymer behavior in the flow plays an essential role in the suppression of cavitation. Figures~\ref{fig:highT} and \ref{fig:lowT} show that the addition of the polymer blurs the vortices, and the vortex formation is delayed. Cressman {\it et al.}~\cite{cbg2001} experimentally illustrated that this phenomenon is caused by the increase in the elongational viscosity owing to the polymer addition. As the elongational viscosity increases, the vorticity decreases. Thus, the minimum pressure at the center of the vortex increases. As shown in Fig.~\ref{fig:void}, in the case of the LJ fluid, most of the cavitation is associated with the formation of vortices owing to the flow separation, such as that corresponding to separated shear layer and recirculation region. In contrast, the presence of polymers reduces the vorticity in the separated shear layer and recirculation region. Consequently, cavitation occurs at a further location where the vortex is formed and emitted. The backward shift of the cavitation location owing to polymers addition has also been observed in previous studies on the step nozzle~\cite{ktk2018, nkm2018}. Therefore, the effects of the polymers on vortex formation contribute to the cavitation suppression.

As shown in Figs.~\ref{fig:highT}, \ref{fig:lowT}, and \ref{fig:map}, the polymer addition changes the thermodynamic state of the fluid and the behavior of the vortices. The LJ fluid and polymer solution exhibit the same tendency for the density field: the density is lower in the region with a large vorticity. However, the temperature field exhibits different trends for the LJ fluid and polymer solution. In the polymer solution, the fluid temperature is higher in the separated shear layer and vortex street behind the cylinder. As shown in Fig.~\ref{fig:polymer}, the polymer is linearly stretched near the cylinder and entrained in the vortex street behind the cylinder. The temperature increase can be attributed to this polymer stretching. Note that this stretching is not caused by the stretching of the bonds connecting the polymer beads, because the mechanical energy of the bonds is almost constant, as shown in Fig.~\ref{fig:polymer}. Thus, it is presumably attributed to the entropic elastic energy due to the polymer deformation into a linear shape by the shear.

As shown in Fig.~\ref{fig:polymer}, the change in the polymer conformation in passing around the cylinder exhibits similar quantitative characteristics in the cases with and without cavitation. Therefore, the polymers exhibit a similar behavior in the flow independent of bubbles. Under both the considered temperatures, the presence of polymers increases the local temperature in the lower fluid density region, as discussed. This behavior of the polymers significantly changes the thermodynamic state of the fluid under a low temperature, as shown in Fig.~\ref{fig:map}. The region below the critical point disappears.  This local temperature increase results in a dramatic suppression of cavitation.
\section{Summary}\label{summary}
The effects of polymer addition on the cavitating flow around a circular cylinder were investigated by performing large-scale MD simulations. The polymer addition significantly suppresses the cavitation. Although the polymer has negligible effects on the gas--liquid phase equilibria, the polymer behavior in the flow impacts the cavitation onset. We clarified that the local temperature increase is essential for cavitation suppression. The flow-induced polymer conformational change increases the local temperature. Thus, the microscopic changes in the polymer chains can affect the macroscopic behavior.

Although we conduct an extremely large-scale MD simulation, the system size is extremely small compared to the scale of typical experiments. However, we believe that our simulation captures essential aspects and the molecular scale insights gained here are useful in a deeper understanding of the experimental facts.
MD simulations do not require a phenomenological model for phase transition and polymer motion. In key engineering flows, the difference in the micro-scale structures often exerts a significant impact on the macro-scale flow. The MD simulation is a suitable tool to investigate such micro-scale effects and can thus facilitate the understanding of multiphase flows in complex fluids.
\section*{DATA AVAILABILITY}
The data that support the findings of this study are available from the corresponding author upon reasonable request.

\begin{acknowledgments}
  This research was supported by MEXT as “Exploratory Challenge on Post-K computer” (Challenge of Basic Science—Exploring Extremes through Multi-Physics and Multi-Scale Simulations) and JSPS KAKENHI, Grant Nos. JP15K05201 and 19H05718. 
We acknowledge the Supercomputer Center, Institute for Solid State Physics (ISSP), University of Tokyo; the Research Center for Computational Science (RCCS), Okazaki, Japan; and the Center for Computational Materials Science, Institute for Materials Research (IMR), Tohoku University, for the use of their supercomputers (Project No.20S0026 for MASAMUNE-IMR in IMR).
\end{acknowledgments}
%
%

%
%

%


\appendix*

\section{Calculation Method of Viscosity}

The viscosity $\eta$ is calculated using the momentum exchange method~\cite{muller99}. The systems for the polymer solution and LJ fluid are rectangular parallelepipeds with $L_x \times L_y \times L_z = 500 \times 100 \times 500$ and $L_x \times L_y \times L_z = 100 \times 100 \times 120$, respectively. The systems are divided into $50$ slabs and $24$ slabs along the $z$ axis for the polymer solution and LJ fluid, respectively. The velocity gradient is generated by exchanging the momentum in the $x$-direction of the particles in the first and middle slabs ($26$th and $13$th slabs for the polymer solution and LJ fluid, respectively). When the total momentum exchanged during the time $t_{\rm sim}$ is $P_x$, the momentum flux $J_z$ in the $z$-direction is given by $J_z=P_x / (2 t_{\rm sim} L_x L_y)$. Since two shear flows with positive and negative gradients are generated in the lower and upper regions, respectively, factor $2$ in the denominator appears. The viscosity coefficient is calculated from 
the relationship between the velocity gradient and momentum flux,
\begin{eqnarray}
  J_z=-\eta \frac{{\rm d}v_x}{{\rm d}z}.
\end{eqnarray}

For this shear simulation, we first create an equilibrium state with temperature $T=2$. MD simulations are performed under a constant number of particles, volume, and temperature. The temperature is kept constant by using a Langevin thermostat. Subsequently, the Langevin thermostat is removed, and momentum exchange is initiated. Note that the temperature decrease during the shear measurement is very small ($\Delta T < 0.1$).
As shown in Fig.~\ref{fig:viscosity}, the polymer solution shows the shear-thinning behavior for shear rates up to $\dot{\gamma}\simeq0.002$. However, at higher shear rates, $\dot{\gamma} \gtrsim 0.006$, the polymer and solvent separated due to the finite size effects, and therefore we could not estimate the viscosity correctly. The polymer chains form two bands and low-polymer-concentration layers appear around the first and middle slabs.  Thus, we used extrapolation to estimate the viscosity at higher shear rates.
%
%
\end{document}